\documentclass{article}
\usepackage{spconf,amsmath,graphicx}

\usepackage{amsmath,amssymb,color,amsmath, graphicx, float, caption, subcaption, mathrsfs,color} \usepackage[ruled]{algorithm2e}
\usepackage{bm}
\usepackage{epstopdf }
\usepackage{cite}

\usepackage{scalerel,stackengine}
\stackMath
\newcommand\reallywidehat[1]{%
\savestack{\tmpbox}{\stretchto{%
  \scaleto{%
    \scalerel*[\widthof{\ensuremath{#1}}]{\kern-.6pt\bigwedge\kern-.6pt}%
    {\rule[-\textheight/2]{1ex}{\textheight}}
  }{\textheight}%
}{0.5ex}}%
\stackon[1pt]{#1}{\tmpbox}%
}
\usepackage{selinput}
\SelectInputMappings{%
  eacute={é},
}

\def\A{\bm{A}}

\def\N{\bm{N}}

\def\U{\bm{U}}
\def\V{\bm{V}}

\def\Z{\bm{Z}}

\def\a{\mathbf{a}}

\def\n{\bm{n}}

\def\n{\bm{n}}

\def\s{\bm{s}}

\def\x{\bm{x}}

\def\bLambda{\boldsymbol{\Lambda}}

\def\balpha{\boldsymbol{\alpha}}

\def\btheta{\boldsymbol{\theta}}

\def\minim#1{\underset{#1}{\textrm{min}}}

\def\CRLB_pe{\text{CRLB}}
\def\CRLB_pe{\text{CRLB}(\mathbf{p}_{e})}

\def\N02{\frac{N_0}{2}}

\def\0{\mathbf{0}}
\def\1{\mathbf{1}}

\SetKwInput{kwEvaluate}{Evaluate}
\SetKwInput{kwSort}{Sort}
\SetKwInput{kwInput}{Input}
\SetKwInput{kwOutput}{Output}
\SetKwInput{kwInitialize}{Initialize}
\SetKwInput{kwParameters}{Params.}
\SetKwInput{kwDefaults}{Default init.}

\title{Direction of Arrival Estimation for Non-Coherent Sub-Arrays via 
Joint Sparse and Low-Rank Signal Recovery}
\name{ Tom~Tirer $^{1,2}$ and Oded~Bialer $^1$
}
\address{$^1$General Motors - Advanced Technical Center Israel\\
$^2$School of Electrical Engineering, Tel Aviv University, Tel Aviv, Israel
}

\begin{document}

\maketitle


\begin{abstract}

Estimating the directions of arrival (DOAs) of multiple sources from a single snapshot obtained by a coherent antenna array is a well-known problem, 
which can be 
addressed by sparse signal reconstruction methods, where the DOAs are estimated from  
the peaks of the recovered high-dimensional signal.
In this paper, we consider a more challenging DOA estimation task where the array is composed of non-coherent sub-arrays (i.e., sub-arrays that observe different unknown phase shifts due to using low-cost unsynchronized local oscillators). 
We formulate this problem as the reconstruction of a joint sparse and low-rank matrix and solve its 
convex relaxation. 
While the DOAs can be estimated from the solution of the convex problem, 
we further show how an improvement is obtained if instead 
one estimates from this solution the phase shifts, creates ``phase-corrected" observations and applies another final 
(plain, coherent) sparsity-based DOA estimation. 
Numerical experiments show 
that the proposed approach outperforms strategies that are based on non-coherent processing of the sub-arrays as well as other sparsity-based methods.

\end{abstract}



\section{Introduction}
\label{INTRO}

Estimating the directions of arrival (DOAs) of radio frequency sources using an array of antenna elements 
is an important problem that 
attracts much interest in different disciplines, such as signal processing and communications 
\cite{krim1996two}.
In many practical applications, e.g.~when the sources and/or receivers are moving, it is required to estimate the DOAs from a single realization (snapshot) of the measurements. 
Several existing strategies can handle the single snapshot scenario. 
While the classical methods are based on minimization (often greedy) of the negative log-likelihood function \cite{ziskind1988maximum, bresler1986exact, feder1988parameter} and 
more recent 
methods are based on training deep neural networks \cite{bialer2019performance}, another popular approach is based on formulating the problem as a sparse signal reconstruction task, and estimating the DOAs from the peaks of the magnitude of the recovered high-dimensional signal \cite{jeffs1998sparse, malioutov2005sparse,yang2018sparse}.

When the antenna elements are {\em coherent},\footnote{Coherent elements are elements that are connected to the same local oscillator. Therefore, ignoring the noise, the phase differences between their observations are only due to the difference in the propagation delays between the location of the sources and the location of the elements.}
the DOA estimation performance improves with the increase of the array aperture 
\cite{van2004optimum, stoica1988music}.
However, maintaining the coherence of all the elements 
of a large array 
is very demanding, especially at high carrier frequency, as it requires expensive hardware and a complicated calibration process that is prone to errors.
Therefore, 
a possible alternative, with lower hardware complexity and cost, 
is to split a large array into several smaller sub-arrays, such that the coherence of the elements is kept only within each sub-array, while the large aperture of the entire array is exploited using signal processing techniques \cite{tirer2020method}. 

Most 
of the methods that can be applied to estimate the DOAs for non-coherent sub-arrays are either based on non-coherent processing of the sub-arrays  
(handling the sub-arrays 
as if they observe different signals) \cite{rieken2004generalizing, wen2014improved, suleiman2018non}, or require a large number of snapshots with the same phase offsets between the sub-arrays (note, though, that unsynchronized
local oscillators lead to different phase offsets in each snapshot)  
\cite{pesavento2001direction, see2004direction, parvazi2011direction, friedlander1991direction, swindlehurst2001exploiting}.
Therefore, to advance the performance of DOA estimation from a single snapshot of non-coherent sub-arrays, the work in \cite{tirer2020method} has proposed an approximate maximum likelihood approach.
However, this approach 
assumes that the number of sources is known. Moreover, it 
is based on non-convex optimization with respect to an optimization variable whose dimension is the number of sources. Therefore, it requires a multidimensional search, or, when solved using iterative schemes, it 
depends on the quality of the initialization.

In this paper, we propose a novel method to estimate the DOAs from a single snapshot of non-coherent sub-arrays. It is based on convex optimization, and thus does not rely on good initializations and can easily handle scenarios with a large (possibly unknown) number of sources. 
Borrowing ideas from sparsity-based DOA estimation for coherent arrays \cite{malioutov2005sparse} and from low-rank-based blind deconvolution \cite{ahmed2013blind}, we formulate the problem as the reconstruction of a joint sparse and low-rank matrix and solve its 
convex relaxation. 
Then, the DOAs can be estimated from the solution of the convex problem.
Yet, we show how an improvement is obtained if instead of estimating the DOAs, this solution is used to estimate the sub-arrays' phase shifts, 
create a ``phase-corrected" observation vector and use it to estimate the DOAs via another final (plain, coherent) sparsity-based DOA estimation.
Numerical experiments show 
that the proposed approach outperforms other 
spectrum-based strategies, 
such as a variant of MUSIC \cite{schmidt1986multiple} and a sparsity-based method that adapts self-calibration algorithms \cite{ling2015self,hung2017low} to the considered problem.


\section{Problem Formulation}
\label{Sec2}

Consider $Q$ far-field sources, located at unknown angles $\btheta \triangleq [\theta_1, \ldots, \theta_Q]^T$ and transmitting unknown narrowband signals that impinge on an $M$-element receiving array. 
Let the array be composed of $L$ sub-arrays, where the $\ell$th sub-array consists of $M_\ell$ elements 
($\sum \limits_{\ell=1}^{L} M_\ell = M$).
We assume that each sub-array is coherent, while different sub-arrays are non-coherent, i.e. 
there is an unknown phase shift between sub-arrays. 
Furthermore, we assume that only a single snapshot is available.

The waveform observed by each of the $L$ sub-arrays can be formulated as
\begin{align}
\label{Eq_subarray_model}
\x_{\ell} = \mathrm{e}^{-j \phi_{\ell}} \tilde{\A}_{\ell}(\btheta)\tilde{\s} + \n_{\ell}, \,\,\,\, \ell=1,\ldots, L,
\end{align}
where $\phi_{\ell}$ is the unknown phase in the $\ell$th sub-array, 
$\tilde{\A}_{\ell}(\btheta) \triangleq [\a_{\ell}(\theta_1), \ldots, \a_{\ell}(\theta_Q)] \in \mathbb{C}^{M_\ell \times Q}$ 
is the $\ell$th sub-array manifold, where each column $\a_{\ell}(\theta)$ is the $\ell$th sub-array response to a signal which arrives at angle $\theta$. 
The vector 
$\tilde{\s} \in \mathbb{C}^{Q \times 1}$ 
represents the unknown signals, and 
$\n_{\ell} \in \mathbb{C}^{M_\ell \times 1}$ 
represents white, zero-mean, circular complex Gaussian noise. 
To simplify the formulation, we assume that the noise variance $\sigma^2$ is known, and equal at all receivers (the extension to
non equal variances is straightforward).
Regarding the steering vector $\a_{\ell}(\theta)$, 
assuming that the axis origin (i.e. the reference point) is defined in the middle of the entire array, and that the source angle is measured with respect to the boresight, then 
\begin{align}
\label{Eq_subarray_response}
[\a_{\ell}(\theta)]_i = g_{\ell,i}(\theta) \mathrm{e}^{j \frac{2\pi}{\lambda}(x_{\ell,i}\mathrm{sin}\theta + y_{\ell,i}\mathrm{cos}\theta)},
\end{align}
where $\lambda$ is the signal wavelength, $(x_{\ell,i},y_{\ell,i})$ denotes the antenna location, and $g_{\ell,i}$ denotes the antenna element pattern, which is assumed to be known. 
For example, for linear arrays of omnidirectional sensors we have that 
$g_{\ell,i}(\theta)=1$ and $y_{\ell,i}=0$.

To conclude this section, 
the problem at hand is to estimate the DOAs $\btheta$ from the sub-arrays' observations $\{ \x_{\ell} \}$ given in \eqref{Eq_subarray_model}, where $\{\phi_{\ell}\}$ and $\tilde{\s}$ are nuisance parameters.


\section{The Proposed Method}
\label{Sec3}

We propose to look at the problem from the sparse signal reconstruction  perspective \cite{malioutov2005sparse}. Namely, we define a DOA grid of length $N_\theta \gg Q$ (containing different hypotheses of a  scalar DOA) and for each sub-array create the 
measurement matrix  $\A_{\ell} \triangleq [\a_{\ell}(\theta_1), \ldots, \a_{\ell}(\theta_{N_\theta})] \in \mathbb{C}^{M_\ell \times N_\theta}$.
The observation model in \eqref{Eq_subarray_model} can now be reformulated as
\begin{align}
\label{Eq_subarray_model_sparse}
\x_{\ell} = \alpha_{\ell}^* \A_{\ell} \s + \n_{\ell}, \,\,\,\, \ell=1,\ldots, L,
\end{align}
where $\alpha_{\ell} \triangleq \mathrm{e}^{j \phi_{\ell}}$ (the superscript $^*$ denotes the complex conjugate), and 
$\s \in \mathbb{C}^{N_\theta \times 1}$ 
obeys $\|\s\|_0=Q$, where $\|\cdot\|_0$ is the $\ell_0$ pseudo-norm that counts the number of non-zero entries of the vector.
Note that the complicated non-linear model in \eqref{Eq_subarray_model} is replaced with the model in \eqref{Eq_subarray_model_sparse} that is bilinear with respect to the unknown parameters $\balpha \triangleq [\alpha_{1}, \ldots, \alpha_{L}]^T$ and $\s$. This suggests that, similarly to the ``lifting" strategy used in \cite{ahmed2013blind} for the blind deconvolution problem (which is also bilinear), we can express \eqref{Eq_subarray_model_sparse} as a linear model with respect to a rank-1 matrix $\Z \triangleq \s \balpha^H  \in \mathbb{C}^{N_\theta \times L}$. Indeed, 
\begin{align}
\label{Eq_subarray_model_sparse_rank1}
\x_{\ell} = \A_{\ell} \Z[:,\ell] + \n_{\ell}, \,\,\,\, \ell=1,\ldots, L,
\end{align}
where $\Z[:,\ell]$ denotes the $\ell$th column of $\Z$.
Note that $\Z$ is not only rank-1, but also ``row-sparse", i.e., only $Q$ of its rows are non-zero. 
Note also that, in practice, the number of sources, $Q$, may not be known in advance.
Therefore, a natural non-convex optimization problem that arises from \eqref{Eq_subarray_model_sparse_rank1} is
\begin{align}
\label{Eq_opt_prob_nonconvex}
&\minim{\Z} \,\,\,  \|\Z\|_{0,2} + \mu \cdot \mathrm{rank}(\Z) \nonumber\\
&\mathrm{s.t.} \,\,\,\, \sum \limits_{\ell=1}^{L} \left \|\x_{\ell} - \A_{\ell} \Z[:,\ell] \right \|_2^2 \leq C M \sigma^2,
\end{align}
where $\mu$ and $C$ are positive parameters, $\|\cdot\|_2$ stands for the Euclidean norm, and $\|\cdot\|_{0,2}$ is a pseudo-norm that counts the number of non-zero rows of the matrix (or equivalently, counts the number of rows whose Euclidean norm is non-zero).
Note that we currently ignore the fact that $\{\alpha_\ell\}$ have unit magnitude. Yet, this information will be used later on.

Hypothetically, 
given a solution to \eqref{Eq_opt_prob_nonconvex}, one can extract from it an estimator of the signal $\s$, and estimate the DOAs as the angles associated with indices of the peaks of the magnitude of the latter estimator. 
However, \eqref{Eq_opt_prob_nonconvex} is a non-convex problem (due to its objective) that is not tractable to solve.   
Therefore, 
to obtain a convex problem that can be efficiently solved we perform convex relaxation on \eqref{Eq_opt_prob_nonconvex}. Following the common practice \cite{chen1994basis,fazel2003matrix}, we replace $\mathrm{rank}(\Z)$ with the nuclear norm $\|\Z\|_{*}$ that sums the singular values of the matrix, and replace $\ell_0$-type pseudo-norms with their $\ell_1$-norm counterparts, i.e., we replace $\|\Z\|_{0,2}$ with $\|\Z\|_{1,2} \triangleq \sum \limits_{n=1}^{N_\theta} \sqrt{ \sum \limits_{\ell=1}^{L} |Z[n,\ell]|^2 }$.
Therefore, we get the following convex optimization problem
\begin{align}
\label{Eq_opt_prob_convex}
&\minim{\Z} \,\,\, \|\Z\|_{1,2} + \mu \|\Z\|_* \nonumber\\
&\mathrm{s.t.} \,\,\,\, \sum \limits_{\ell=1}^L \left \|\x_{\ell} - \A_{\ell} \Z[:,\ell] \right \|_2^2 \leq C M \sigma^2.
\end{align}
This convex program 
can be directly solved by the CVX Matlab package \cite{grant2011cvx} (that applies the SDPT3 solver \cite{toh1999sdpt3}).
Since typically $L$ (the number of sub-arrays) is small, this procedure has tractable runtime  (it took only a few seconds in our experiments with Intel-i7 laptop).

Let us denote by $\hat{\Z}_{\mathrm{cvx}}$ the solution to \eqref{Eq_opt_prob_convex}. 
Although the nuclear norm promotes low-rankness, it does not strictly impose rank-1. To strictly impose it, 
we apply 
the singular value decomposition (SVD) on $\hat{\Z}_{\mathrm{cvx}}$.
Namely, we decompose $\hat{\Z}_{\mathrm{cvx}} =\U \bLambda \V^H$, where $\U \in \mathbb{C}^{N_\theta \times N_\theta}$ and $\V \in \mathbb{C}^{L \times L}$ are unitary matrices and $\bLambda$ is a rectangular diagonal matrix.
Now, the rank-1 approximation of $\hat{\Z}_{\mathrm{cvx}}$ is given by $\hat{\Z}=\Lambda[1,1]{\U[:,1]}\V[:,1]^H$, where $\hat{\s}=\Lambda[1,1]\U[:,1]$ and $\hat{\balpha}=\V[:,1]$ can be used as proxies to $\s$ and $\balpha$.

We turn to develop {\em two} DOA estimation strategies. 
The first strategy, which has already been mentioned above, is to estimate the DOAs as the peaks of the magnitude of the $N_\theta \times 1$ vector $\hat{\s}$. 
Yet, recall 
that until now we ignored the fact that the ``true" $\{\alpha_\ell\}$ have unit magnitude 
(since 
$\alpha_{\ell} = \mathrm{e}^{j \phi_{\ell}}$). To exploit this information, we propose 
another 
DOA estimation strategy 
(which improves the results in our experiments).

First, 
we estimate (only) the phase shifts $\{\phi_{\ell}\}$ from $\hat{\Z}$ by 
\begin{align}
\label{Eq_phase_est}
\hat{\phi}_{\ell} = \angle \hat{\alpha}[\ell] = \angle V[\ell,1], \,\,\,\, \ell=1,\ldots, L,
\end{align}
where $\angle x$ stands for the phase of the complex number $x$. 
Looking back on
the model in \eqref{Eq_subarray_model}, we use the estimated phase shifts to obtain the ``phase-corrected" observations of the entire array, denoted by $\hat{\overline{\x}}$, that can be (approximately) modeled as
\begin{align}
\label{Eq_array_model_phase_corrected}
\hat{\overline{\x}} \triangleq 
\begin{bmatrix}
\mathrm{e}^{j \hat{\phi}_{1}}\x_{1} \\ \vdots \\ \mathrm{e}^{j \hat{\phi}_{L}}\x_{L} 
\end{bmatrix}
\approx \tilde{\A}(\btheta) \tilde{\s} + \overline{\n},
\end{align} 
where $\tilde{\A}(\btheta) \triangleq \left [ \tilde{\A}_1^T(\btheta) \ldots \tilde{\A}_L^T(\btheta) \right ]^T \in \mathbb{C}^{M \times Q}$  
and $\overline{\n} \in \mathbb{C}^M$ is a noise vector, whose statistics are not expected to dramatically change (compared to $\{\n_\ell\}$) due to the phase shifts. 
As a minor remark, note that only the differences between the phase shifts estimates $\{\hat{\phi}_{\ell}\}$ matter, as 
the mean of the estimated phases can be absorbed 
into the unknown vector $\tilde{\s}$.

At this point, any DOA estimation method for {\em coherent array} that can handle the single snapshot case 
can be applied on the ``phase-corrected" $\hat{\overline{\x}}$, modeled in \eqref{Eq_array_model_phase_corrected}.
We choose to use the popular sparsity-based technique from \cite{malioutov2005sparse}. Using the same notations of $\s$ and $\A_\ell$ that are used in \eqref{Eq_subarray_model_sparse}, we reformulate \eqref{Eq_array_model_phase_corrected} as
\begin{align}
\label{Eq_array_model_phase_corrected_sparse}
\hat{\overline{\x}} \approx \A \s + \overline{\n},
\end{align} 
where $\A \triangleq \left [ \A_1^T \ldots \A_L^T \right ]^T  \in \mathbb{C}^{M \times N_\theta}$, 
and estimate $\s$ as the minimizer of the following convex program
\begin{align}
\label{Eq_opt_prob_convex_phase_corrected}
&\minim{\s} \,\,\, \|\s\|_{1} \nonumber\\
&\mathrm{s.t.} \,\,\,\, \left \| \hat{\overline{\x}} - \A \s \right \|_2^2 \leq C M \sigma^2,
\end{align}
where $C$ is a positive parameter and $\|\cdot\|_1$ is the $\ell_1$-norm.
Again, this problem can be solved fastly using CVX \cite{grant2011cvx}, and we denote by $\hat{\s}_{\mathrm{cvx}}$ this solution. 
Finally, as done in \cite{malioutov2005sparse}, we estimate the DOAs as the peaks of the magnitude of the $N_\theta \times 1$ vector $\hat{\s}_{\mathrm{cvx}}$.


\section{Numerical Results}
\label{Sec4:numerical}

We perform 
computer simulations 
to evaluate the performance of the two proposed DOA estimation methods.
The strategy where we estimate the DOAs from $\hat{\Z}$ will be referred to as {\em Proposed1}.
The strategy where we estimate the phase shifts from $\hat{\Z}$ and then apply (quasi-)coherent sparsity-based method will be referred to as {\em Proposed2}.
In all the experiments we use the parameters $\mu=1$ and $C=2$ in \eqref{Eq_opt_prob_convex}. 

We compare our strategies with two reference methods.
The first reference method is a sparsity-based approach that adapts the self-calibration algorithms \cite{ling2015self,hung2017low} to the considered non-coherent problem. Specifically, this technique obtains a matrix $\tilde{\Z} \in \mathbb{C}^{N_\theta \times L}$ using a convex program that is similar to \eqref{Eq_opt_prob_convex} except that its objective includes only $\|\Z\|_{1,2}$ (i.e., $\mu=0$ and $C=2$). Then, following \cite{hung2017low} (that has presented better results than \cite{ling2015self}), the rank-1 approximation of $\tilde{\Z}$ is obtained using SVD. The DOAs are estimated as the peaks of the magnitude of the $N_\theta \times 1$ 
most dominant left singular vector of $\tilde{\Z}$. 
This method will be referred to as {\em SparsityOnly}.
Observe that the difference between our {\em Proposed1} and {\em SparsityOnly} is the fact that we include the nuclear norm in the convex optimization.

The second reference method is based on MUSIC \cite{schmidt1986multiple}. 
It 
treats the realizations of the $L$ sub-arrays as $L$ snapshots of a single sub-array. We use this method, which will be referred to as {\em MUSIC}, with spatial smoothing \cite{shan1985spatial} and forward-backward smoothing \cite{pillai1989forward} (within each sub-array) that empirically improved its performance. Applying this non-coherent MUSIC is possible since in our experiments the sub-arrays are uniform linear arrays  (ULAs) 
with similar sensor configuration. 
Note that all the examined methods are based on estimating the DOAs as the peaks of $N_\theta \times 1$ ``spectrums", 
and  
that only {\em MUSIC} requires the number of sources, $Q$, 
for 
generating its spectrum.

We consider 
a ULA of $M=24$ elements, with half wavelength spacing, partitioned into $L=4$ sub-arrays of size $M_{\ell}=6$. 
We start with $Q=2$ 
sources 
located at angles 
$\{ 0^\circ, 15^\circ \}$
(the boresight angle of the entire array is $0^\circ$)  
that  
transmit complex Gaussian signals with equal power.
The unknown phase of each sub-array is selected at random from a uniform distribution over $[0, 2\pi]$. 
To obtain statistical results, we perform $N_{\text{exp}}=250$ Monte Carlo experiments at each SNR between 0 dB to 30 dB. In each experiment we draw the Gaussian noise $\{\n_{\ell}\}$, the signal $\tilde{\s}$, and the unknown phases $\{\phi_{\ell}\}$.
For each method, the performance is measured by the root mean square error (RMSE) of the $Q$ highest peaks.

\begin{figure}
  \centering
  \begin{subfigure}[b]{0.9\linewidth}
    \centering\includegraphics[width=150pt]{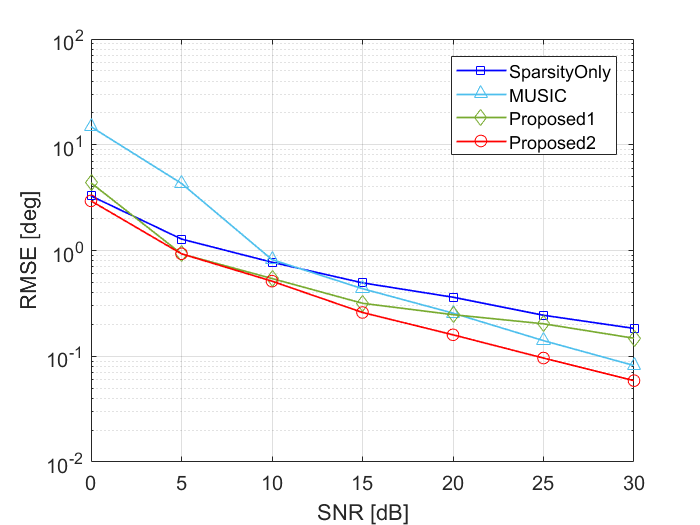}
  \end{subfigure}%
  \caption{RMSE vs. SNR for sources at $0^\circ$ and $15^\circ$, using 4 uniform linear sub-arrays, each of 6 elements.}
\label{fig:Q4_L4_M24_new2}

\vspace{5mm}

  \centering
  \begin{subfigure}[b]{0.9\linewidth}
    \centering\includegraphics[width=150pt]{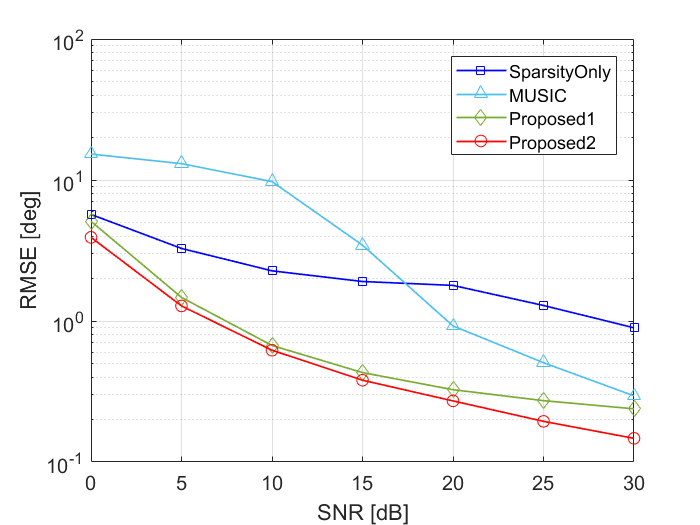}
  \end{subfigure}%
  \caption{RMSE vs. SNR for sources at $-15^\circ$, $0^\circ$, $15^\circ$ and $30^\circ$, using 4 uniform linear sub-arrays, each of 6 elements.}
\label{fig:Q4_L4_M24_new}
\end{figure}

The RMSE results are presented in Fig.~\ref{fig:Q4_L4_M24_new2}. 
It can be seen that {\em Proposed1}, which promotes low-rankness {\em within} the convex optimization, it better than {\em SparsityOnly}.
The best accuracy is observed for our {\em Proposed2}.
The better performance of {\em Proposed2} over {\em Proposed1} (especially at high SNR) demonstrates the advantage of using the solution of the joint sparse and low-rank optimization problem for estimating the phase shifts rather than the DOAs.

\begin{figure}
  \centering
  \begin{subfigure}[b]{0.49\linewidth}
    \centering\includegraphics[width=120pt]{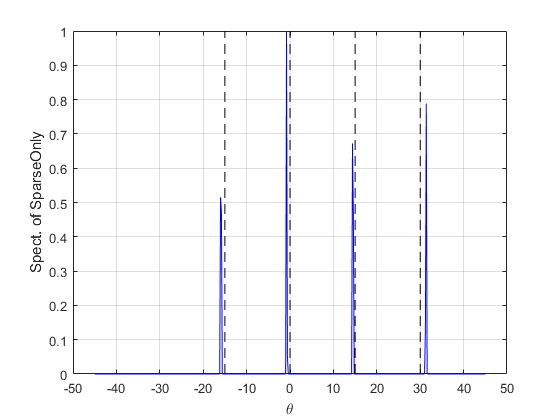}
  \end{subfigure}%
  \begin{subfigure}[b]{0.49\linewidth}
    \centering\includegraphics[width=120pt]{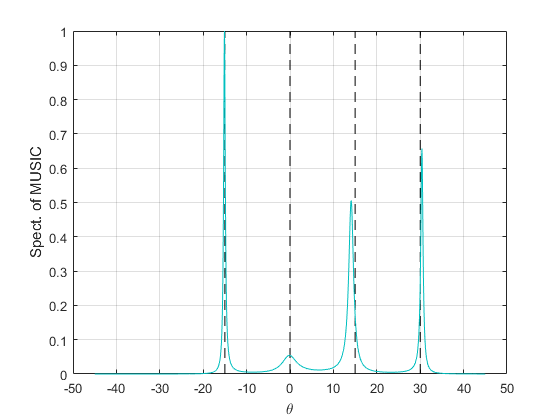}
  \end{subfigure}%
\\
  \begin{subfigure}[b]{0.49\linewidth}
    \centering\includegraphics[width=120pt]{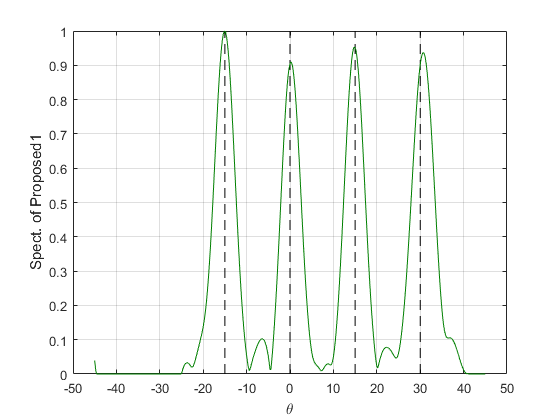}
  \end{subfigure}
  \begin{subfigure}[b]{0.49\linewidth}
    \centering\includegraphics[width=120pt]{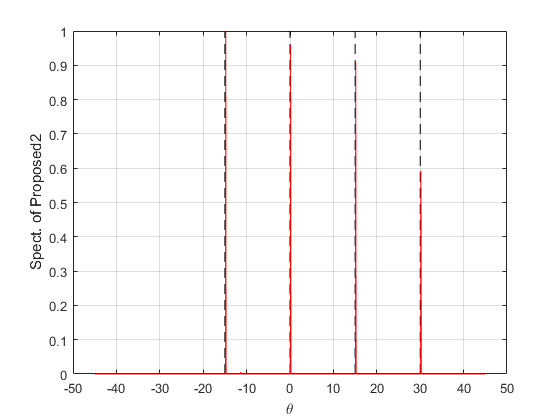}
  \end{subfigure}
  \caption{The spectrums of the different methods for sources at $-15^\circ$, $0^\circ$, $15^\circ$ and $30^\circ$, using 4 ULAs, each of 6 elements, for one realization at SNR of 20 dB. From left to right and top to bottom: {\em SparsityOnly}, {\em MUSIC}, {\em Proposed1}, and {\em Proposed2}.}
\label{fig:spectrums}
\end{figure}

Spectrum-based methods are especially preferable 
over maximum likelihood techniques 
when the number of sources is large. 
Thus, we turn to examine a scenario with $Q=4$ sources, located at angles $\{ -15^\circ, 0^\circ, 15^\circ, 30^\circ \}$. The rest of the configuration remains as before. The RMSE results are presented in Fig.~\ref{fig:Q4_L4_M24_new}, and the estimators' spectrums for one realization at SNR of 20 dB are displayed in Fig.~\ref{fig:spectrums}. 
Comparing the spectrums of {\em Proposed1} and {\em SparsityOnly} in Fig.~\ref{fig:spectrums}, we see that the nuclear norm that is used in {\em Proposed1} reduces the sparsity but significantly improves the accuracy of the DOAs estimates (i.e., the location of the peaks).
We further see from Fig.~\ref{fig:spectrums}
that {\em MUSIC}'s peaks are quite accurate for this realization, though, one source is almost not identified. Such detection failures explain the bad results of {\em MUSIC} at low to medium SNR in Fig.~\ref{fig:Q4_L4_M24_new}.
Clearly, the best accuracy is observed for our {\em Proposed2}.

Note that in Fig.~\ref{fig:Q4_L4_M24_new} the advantage of the two proposed methods over the reference methods is larger than in Fig.~\ref{fig:Q4_L4_M24_new2}.
This can be explained by the property that having more sources improves the phase offsets estimation (see the analysis in \cite{tirer2020method}).
Specifically, this property directly explains the improved results compared to {\em MUSIC} (which is a non-coherent processing method).
As for the improved advantage over {\em SparsityOnly}, note that the phase shifts estimation relates to the nuclear norm term in \eqref{Eq_opt_prob_convex}.
While this term is used for both {\em Proposed1} and {\em Proposed2}, it is eliminated ($\mu=0$) for {\em SparsityOnly}.

\section{Conclusions}
\label{Sec5:conc}

We addressed the problem of estimating the DOAs of multiple sources from a single snapshot of non-coherent sub-arrays.
We formulated this problem as the reconstruction of a joint sparse and low-rank matrix and solved its convex relaxation. Unlike methods that are based on maximum likelihood \cite{tirer2020method}, our approach 
can easily handle scenarios with a large (and possibly unknown) number of sources. 
We showed two ways to estimate the DOAs. The first strategy, estimates the DOAs directly from the solution of the convex problem and outperforms an existing sparsity-based method that does not promote low-rankness within the optimization.
The second strategy, which has shown even improved results, uses the solution of the convex problem for estimating the phase shifts rather than the DOAs. The estimated phase shifts are then used to ``correct" the observations, which allows DOA estimation with plain methods of coherent arrays.
We note that there 
is 
work that questions the benefit of combining nuclear and $\ell_1$ norms for exact reconstruction of an entire high-dimensional matrix \cite{oymak2015simultaneously}. 
However, our paper shows that when the goal is to estimate from the observations only a  
rather small set
of fundamental parameters (e.g., DOAs and phase shifts) then this combination of norms 
can be 
beneficial.



\bibliographystyle{ieeetr}
\bibliography{main}

\end{document}